\documentstyle[12pt,epsfig]{ioplppt}

\eqnobysec

\newcommand{\beq}{\begin{equation}}
\newcommand{\eeq}{\end{equation}}
\newcommand{\eeql}[1]{\label{#1}\end{equation}}

\newcommand{\bea}{\begin{eqnarray}}
\newcommand{\eea}{\end{eqnarray}}
\newcommand{\eeal}[1]{\label{#1}\end{eqnarray}}

\newcommand{\nid}{\noindent}
\newcommand{\nn}{\nonumber \\}
\newcommand{\sfrac}[2]{\mbox{\small $\frac{#1}{#2}$}} 
\newcommand{\bfrac}[2]{\frac{\displaystyle #1}{\displaystyle #2}}

\newcommand{\norm}[1]{\langle #1 | #1 \rangle} 
\newcommand{\me}[3]{\langle #1 | #2 | #3 \rangle} 

\begin{document}
\jl{1}

\title{Logarithmic perturbation theory for quasinormal modes}

\author{P T Leung\dag, Y T Liu\dag, 
W M Suen\dag\ddag, C Y Tam\dag\ and K Young\dag}

\address{\dag\ Department of Physics, 
The Chinese University of Hong Kong, Hong Kong, China}
\address{\ddag\ MacDonnel Center for the Space Sciences, Department of 
Physics, Washington University, St Louis, Missouri 63130, USA}

\date{\today}
 
\begin{abstract}
Logarithmic perturbation theory (LPT) is developed and applied to quasinormal
modes (QNMs) in open systems.  QNMs often do not form a complete
set, so LPT is especially convenient because summation over a
complete set of unperturbed states is not required.
Attention is paid to potentials with exponential tails,
and the example of a P\"{o}schl-Teller potential
is briefly discussed.  A numerical method is developed that handles the 
exponentially large wavefunctions which appear in dealing with QNMs. 
\end{abstract}



\section{Introduction\label{sec:1}}

\subsection{Logarithmic perturbation theory}

Eigenvalue problems of the type 

\beq
H \phi = \lambda \phi
\eeql{eq:eigeneq}

\nid
occur in many branches of physics; here $\lambda$ may be related to the frequency
$\omega$ by $\lambda = \omega$ (Schr\"{o}dinger equation), or by
$\lambda = \omega^2$ (wave equation or Klein-Gordon equation).  Perturbation
theory is useful for systems that depart slightly from an ideal
solvable configuration.  Apart from the standard Rayleigh-Schr\"{o}dinger
perturbation theory (RSPT), a useful alternative 
focuses not on $\phi$ itself, but on its
logarithmic derivative $f = \phi' / \phi$.
Known as logarithmic perturbation theory (LPT) \cite{went,price,polik,aa1},
this method is commonly applied 
to 1-d bound state problems, especially the ground state.  
For excited states, one either has to first factor out the zeros
\cite{zero1}, or
detour around them in the complex plane \cite{zero2}.  
LPT has also been developed for bound
states in 3 d \cite{3d}.
LPT avoids sums over intermediate
states, and comparison with RSPT can lead to
useful sum rules \cite{sum}.

The bound states, or normal modes (NMs), are solutions 
with $\phi \rightarrow 0$
as $x \rightarrow \pm \infty$.  Other boundary conditions are also important
physically.  Scattering states and the phase shift can be handled
by LPT using only on-shell information \cite{phase-shift}.
In this paper, we develop LPT for wavefunctions that are outgoing at infinity
--- quasinormal modes (QNMs).

\subsection{Quasinormal modes\label{sec:1.2}}

In conservative systems, NMs are factorized solutions
$\Phi(x,t) = e^{-i \omega t} \phi(x)$ with $\phi$ satisfying
an eigenvalue equation such as (\ref{eq:eigeneq}) and nodal boundary 
conditions at $x \rightarrow \pm \infty$.
The counterparts in open systems are QNMs; these factorized solutions 
satisfy outgoing wave boundary conditions at $x \rightarrow \pm \infty$, 
so that ${\rm Im}\, \omega \equiv - \gamma < 0$.

QNMs are important from many points of view.
A laser is often discussed in terms of its ``modes'', i.e., 
the spectral lines with finite widths $\gamma$, which are 
precisely these QNMs \cite{lamb}.  Quantum-mechanical resonances are likewise
central to scattering \cite{a,b,c,gold}, and as intermediate states in high-order transitions.
Gravitational waves from the vicinity of a black hole
are likely to be detected in the next decade by facilities
such as LIGO and VIRGO \cite{detector}. The radial wavefunction 
describing the propagation of gravitational waves in any angular momentum 
sector satisfies a Klein-Gordon equation with a potential $V(x)$ 
\cite{chand,zerilli}.
Theoretical studies \cite{bholen,others} show that,
at least for an intermediate time domain, the waves
are dominated by a ringing signal, which is readily identified as the
superposition of QNMs \cite{bholen,others,ching-tail}.  If the 
relationship between the characteristics of the ringing signal (i.e., the QNMs)
and the spacetime curvature
could be better understood, gravitational waves have the prospect of
becoming a novel astronomical probe.
In these cases, the background is a Schwarzschild metric plus perturbations
(e.g., due to an accretion disk), so perturbative treatments will be useful.

In the present context, three properties of QNMs should be emphasized.  First,
their numerical determination is
notoriously difficult.  This is most simply seen in the
``shooting'' algorithm: choose an arbitrary $\omega$,
integrate from one end (say $x=0$ for a half-line
problem or a full-line problem with definite parity),
identify the coefficient of the
``wrong'' solution at the other end (say $x \rightarrow \infty$),
and vary $\omega$
until this coefficient is zero.  For NMs,
the exponentially large ``wrong'' solution is
readily identified.  For QNMs, the ``wrong'' solution,
which is $O(e^{-2\gamma x})$
relative to the ``right'' solution, is difficult
to extract, especially when $\gamma$ is
large.  The numerical
difficulties make perturbation methods even more relevant 
than would be the case for NMs.

Secondly, RSPT is inapplicable for two reasons. 
Its usual derivation relies on the hermiticity of the system,
which is now lost.  Moreover, because the QNMs are in general not complete,
one cannot sum over intermediate states.  Even in circumstances
where the QNMs turn out to be complete 
\cite{b,c,lly1,lly2,lly3,ching-comp,twocomp1,twocomp2}, a scheme such as LPT 
would still have definite advantages, because it makes no reference
to the higher states with large $\gamma$.

Thirdly, it is readily shown that any QNM can have at most one node
on the real $x$ axis.  Except for the
origin for the odd-parity sector of symmetric potentials,
there is no reason why any root of the
{\em complex} equation $\phi(x) = 0$ should lie on the
real $x$ axis; those cases that do are therefore ``accidental''
in the sense that they occur only for specific values
of the parameters defining the potential $V(x)$ ---
in other words on a set of measure zero in parameter space.
Therefore the nodal problem which plagues LPT
for excited NMs is here absent generally.  

\subsection{Outline of paper}

Section \ref{sec:2} develops LPT for QNMs,
and discusses the generalized norm that
emerges as a result.  The most explicit general form for the
second-order correction, together with an illustrative example,
are given in Section \ref{sec:3},
focusing on those cases where both the original potential
and the perturbation have finite support.
The situation becomes slightly more complicated if the potentials have tails,
and the case of exponential tails is discussed in Section \ref{sec:4}.
A conclusion is given in Section \ref{sec:5}.

\section{Perturbation theory\label{sec:2}}

\subsection{Formalism for the eigenvalue}

We deal with the Klein-Gordon equation:

\beq
\left[ \partial_x^2 - V(x) + \omega^2 \right] \, \phi(x) =0 \;.
\eeql{eq:kgom}

\noindent
The Schr\"odinger equation is included by simply re-labeling
$\omega^2 \mapsto \omega$.
The logarithmic derivative 
$f(x)=\phi'(x) / \phi(x)$ satisfies the Riccati equation

\beq
f'(x) + f^2(x) - V(x) + \omega^2 = 0 \;.
\eeql{eq:ricc}

\nid
We let $f$ denote the logarithmic derivative
corresponding to an eigenvalue, so that it 
satisfies the {\em two} boundary conditions 
$f(x) \rightarrow \pm i \omega$ as $x \rightarrow \pm \infty$.
At a general frequency, however, we can define
similar functions $\phi_{\pm}(\omega,x)$
and their logarithmic derivatives $f_{\pm}(\omega,x)$
as solutions to (\ref{eq:kgom}) and (\ref{eq:ricc}), but 
with each function satisfying
only {\em one} boundary condition, namely
$f_{\pm}(\omega,x) \rightarrow \pm i\omega$
as $x \rightarrow \pm \infty$.  At an eigenvalue,
$f_{-}=f_{+}=f$.

Now let the potential be perturbed

\beq
V(x) = V_0(x) + \mu V_1(x) \;,
\eeql{eq:v1}

\noindent
where $\mu$ is a formal small parameter.  The eigenvalue 
$\omega$ and the function $f$ are both written 
in powers of $\mu$\ftnote{1}{It is a property of LPT that one need focus 
only on one state at a time. Therefore, a label for different QNMs will in 
general be suppressed.}: 

\beq
\omega = \omega_0 + \mu \omega_1 + \mu^2 \omega_2 + \cdots \;,
\eeql{eq:omega}

\beq
f = f_0 + \mu f_1 + \mu^2 f_2 + \cdots \equiv f_0 + g \;,
\eeql{eq:f}

\nid
where $f_0$, assumed known, satisfies the Riccati equation (\ref{eq:ricc})
with the potential $V_0$ and frequency $\omega_0$.

Now divide the real line into three regions $(-\infty, L_-)$, $(L_-,L_+)$ and $(L_+,\infty)$.
If the original potential and its perturbation both have finite support
within the central interval,
then the asymptotic regions are trivial, and the simplest examples will
be of this type.

First consider the central region, and put (\ref{eq:omega}) and (\ref{eq:f})
into (\ref{eq:ricc}).  Upon comparing powers of $\mu$,
one finds 

\beq
f_n' + 2f_0f_n + 2\omega_0 \omega_n = V_n \;,
\eeql{eq:gn}

\nid
for $n = 1, 2, \cdots$, in which $V_1$ is the perturbing potential in (\ref{eq:v1}),
and $V_n$, $n>1$, 
is a shorthand for the following combination of lower-order quantities, to be called 
the effective $n$th order potential

\beq
V_n(x) = -\sum_{i=1}^{n-1} \left[ f_i(x) f_{n-i}(x) + \omega_i \omega_{n-i} \right] \;.
\eeql{eq:vn}

\nid
Using the integrating factor $\exp \! \left[ 2 \int^x dy \, f_0(y) \right]=
\phi_0^2(x)$, one gets
from (\ref{eq:gn})

\beq
 f_n(x) \phi_0^2(x) \mbox{\Huge $|$}_{L_-}^{L_+} =
\int_{L_-}^{L_+} dx\, \left[ V_n(x) - 2 \omega_0 \omega_n \right] 
\phi_0^2(x) \;.
\eeql{eq:int}

We now need to match the central solution to the two asymptotic regions.
Assume that the latter have been solved with outgoing
wave boundary conditions at spatial infinity, and denote the logarithmic derivatives to
be matched as

\beq
D_{\pm}(\omega) = f_{\pm}(\omega,L_{\pm}) \;.
\eeql{eq:logderiv}

\nid
Note that $D_{\pm}$ will contain two types of changes from the
unperturbed case.  First, at fixed
$\omega$, the wavefunction when integrated inwards from $\pm \infty$
will suffer changes because of $V_1(x)$ in the two asymptotic regions;
these are expressed through

\beq
D_{\pm}(\omega) = D_{\pm 0}(\omega) + \mu D_{\pm 1}(\omega) 
+ \mu^2 D_{\pm 2} (\omega) + \cdots \;.
\eeql{eq:dl}

\nid
Secondly, there will be changes because the value of $\omega$ itself shifts
according to (\ref{eq:omega}).
In particular, the exact logarithmic derivative is
$f_{\pm}(\omega,L_{\pm}) = D_{\pm}(\omega)$,
whereas the corresponding unperturbed 
quantity is 
$f_0(L_{\pm}) = D_{\pm 0}(\omega_0)$.
Thus

\beq
g(L_{\pm}) = f_{\pm}(\omega,L_{\pm})- f_0(L_{\pm}) 
= D_{\pm}(\omega) - D_{\pm 0}(\omega_0) \;.
\eeql{eq:gmatch}

\nid
Using (\ref{eq:omega}) and (\ref{eq:dl}) and developing 
the right hand side of (\ref{eq:gmatch}) in powers
of $\mu$, we can find that $f_n$ in (\ref{eq:int}) should be matched to

\beq
f_n(L_{\pm}) =  \omega_n D_{\pm 0}' (\omega_0) + \Delta_{\pm n} \;,
\eeql{eq:gnmatch}

\nid
where $\Delta_{\pm n}$ does not contain $\omega_n$; explicitly but in shorthand

\bea
\Delta_1 &=& D_1 \;, \nn 
\Delta_2 &=& D_2+ \omega_1 D_1'
+ \frac{1}{2} \omega_1^2 D_0''\;,
\eeal{eq:deltan}

\nid
etc.  In the above, the subscripts $\pm$ have been omitted from all quantities, and all
$D_n$ on the right are to be evaluated at $\omega_0$.  In short, one requires a knowledge
of the perturbation in the asymptotic region ($D_n$, $n > 0$), as well as a knowledge
of the unperturbed problem slightly away from the original frequency (derivatives
of $D_0$).

Putting these into (\ref{eq:int}) and collecting terms involving $\omega_n$, one finds
the central result

\beq
\omega_n = \frac{\me{\phi_0}{V_n}{\phi_0}}
{ 2 \omega_0 \norm{\phi_0} } \;,
\eeql{eq:result}

\nid
in which we have introduced the suggestive notation

\bea
\fl \me{\phi_0}{V_n}{\phi_0} &=& \int_{L_-}^{L_+} dx  V_n(x) \phi_0^2(x)
+\left[-\Delta_{+n} \phi_0^2(L_+)+\Delta_{-n} \phi_0^2(L_-)\right] \;,
\label{eq:matelm} \\
\fl \hspace{5mm} \norm{\phi_0} &=& \int_{L_-}^{L_+} dx  \phi_0^2(x)
+\frac{1}{2\omega_0} \left[ D_{+0}' \phi_0^2(L_+) 
-D_{-0}' \phi_0^2(L_-) \right]\;.
\eeal{eq:norm}

\nid
This expresses the $n$th order correction in quadrature in terms of lower-order
quantities (provided the asymptotic regions have been solved to give $\Delta_{\pm n}$
and $D_{\pm 0}'$).

The division into three regions is arbitrary, and the whole expression 
(\ref{eq:result}) must be 
independent of $L_{\pm}$.  Moreover, the numerator and denominator
must be separately independent of $L_{\pm}$, because 
the numerator depends on the perturbation, whereas the denominator
relies only on the unperturbed system; an explicit proof can be constructed
by calculating $\partial \norm{\phi_0} / \partial L_{+}$,
and then using (\ref{eq:ricc}).

Thus, in both (\ref{eq:matelm}) and (\ref{eq:norm}), we can formally
take $L_{\pm} \rightarrow \pm \infty$ and write

\bea
\me{\phi_0}{V_n}{\phi_0} &=& \int_{-\infty}^{\infty} dx  V_n(x) \phi_0^2(x) \;,
\label{eq:matelmf} \\ 
\hspace{5mm} \norm{\phi_0} &=& \int_{-\infty}^{\infty} dx  \phi_0^2(x) \;.
\eeal{eq:normf}

\nid
These formal expressions do not converge;
(\ref{eq:matelm}) and (\ref{eq:norm})
may be regarded as ways of regularizing them.  In Section \ref{sec:4}
we shall discuss various different ways of giving meanings to these
formal integrals.

Evidently, the numerator should be regarded as a generalized matrix element, and the
denominator should be regarded as a generalized norm.  We now develop this interpretation.
The corrections to the eigenfunction will be given in Section \ref{sec:2.3}.

\subsection{Generalized norm and matrix element}

For the simplest case where $V_0(x)$ vanishes
outside the interval $(L_-,L_+)$, the solutions in the two asymptotic
regions are exactly $e^{\pm i\omega x}$, and $D_{\pm 0} (\omega)=\pm i\omega$.
The generalized norm (\ref{eq:norm}) simplifies to

\beq
\norm{\phi_0} =
\int_{L_-}^{L_+}dx\, \phi_0^2(x)+\frac{i}{2\omega_0} \left[  \phi_0^2(L_+) 
+ \phi_0^2(L_-) \right]\;.
\eeql{eq:norm2}

\nid
In this form applicable to potentials without tails, the generalized norm has been introduced
previously both for the wave equation \cite{lly1,perem}, the Schr\"{o}dinger
equation \cite{schrod} and the Klein-Gordon equation \cite{ching-comp}, and its properties
discussed.  It has been shown to be equivalent to another form first given by Zeldovich \cite{zel},
which did not have the surface terms, but instead required a process of regularization \cite{a,b,c}
which is less convenient for actual evaluation (especially numerical evaluation).
The present result, in the more general form (\ref{eq:norm}), is however
applicable to potentials with tails, and examples will be given in 
Section \ref{sec:4}.

We next briefly describe the properties of this generalized norm,
and argue why it deserves to be so named.

First of all, suppose the system parameters
can be tuned so that the leakage of the wavefunction
approaches zero (e.g., if $V_0(x)$ contains a tall barrier on both sides).  Then
$L_{\pm}$ can be chosen so that $\phi_0(L_{\pm}) \approx 0$;
the expression in (\ref{eq:norm}) then contains only the
integral.  Moreover, when the leakage is zero, the frequency is real, and the wavefunction
has a constant phase, which can be chosen to be real; thus $\phi_0^2 = |\phi_0|^2$.
The expression (\ref{eq:norm}) then reduces to the usual (real and positive-definite) norm for a NM.
Because of this property, and because it appears in the denominator in 
(\ref{eq:result}) to scale the wavefunction, it is appropriate to call this 
quantity the generalized norm.

Nevertheless, it has some unusual properties.  (a)  It
involves $\phi_0^2$ rather than $|\phi_0|^2$, and is in general a complex quantity.
(b)  It involves a surface term, though the value of the entire
expression is independent of the choice of $L_{\pm}$.
Thus, it is not a genuine norm, and the term is 
merely a shorthand for ``a bilinear map that appears in the place of the norm
in perturbation formulas such as (\ref{eq:result})''.

It is hardly surprising that perturbative results are expressed in the form
of a matrix element divided by a normalizing factor, as in (\ref{eq:result}), but it
would not have been obvious what the normalizing factor should be.
The point is that for a QNM, the wavefunction behaves as
$\phi_0(x) \approx e^{i\omega_0 x} \propto e^{\gamma_0 x}$
as $x \rightarrow \infty$ and $\gamma_0=-{\rm Im}\, \omega_0$, so that an 
expression such as
(\ref{eq:normf}) (and even more so for the analogous formula with 
$|\phi_0|^2$) 
would be divergent.  An important result of the present paper is that we
give a precise way of normalizing such QNM wavefunctions.

We have already remarked that the generalized norm is not real, and neither is
the (diagonal) matrix element.  Far from being a problem, this is necessary, in that
the result (\ref{eq:result}) gives the corrections to both the real part and the imaginary part
of the frequency.  Thus, despite the formal similarity to the analogous problem for NMs,
the present formalism in fact contains twice the amount of information.

Some of the properties above, in particular the validity without regularization of the simpler form
(\ref{eq:matelmf}), relies on $V_n$ behaving mildly at infinity.  It is therefore
appropriate to demonstrate that if the perturbation $V_1$ has finite support,
then so does all the effective potentials $V_n$ generated via (\ref{eq:vn}).  Consider for simplicity
only the asymptotic interval $(L_+,\infty)$.  Now the exact logarithmic derivative is
$i\omega$, whereas the unperturbed analog is $i\omega_0$.  This then gives

\beq
g^2 = - (\omega - \omega_0)^2 = - \left( \sum_{i=1}^{\infty} \mu^i \omega_i \right)^2\;.
\eeql{eq:gsq}

\nid
The $n$th order term in the above expression then ensures that $V_n$
in (\ref{eq:vn}) exactly vanishes in this region. 

\subsection{Wavefunction\label{sec:2.3}}

To complete the iterative procedure, we also need
the eigenfunctions.  This can be readily obtained by integrating (\ref{eq:int})
to an arbitrary point, and using (\ref{eq:gmatch}) as the boundary condition, which yields

\beq
\fl f_n(x) \phi_0^2(x)=
\left[ \omega_n D_{-0}'(\omega_0) + \Delta_{-n} \right ] \phi_0^2(L_-) 
+\int_{L_-}^x dy \left[ V_n(y) - 2\omega_0 \omega_n \right] 
\phi_0^2(y)\;.
\eeql{eq:function}

\nid
One could write an alternate expression using the boundary condition at $L_+$ and
integrating from the right.  Consistency is guaranteed if $\omega_n$
has been correctly evaluated by (\ref{eq:result}).

Thus we have in principle an order-by-order iteration scheme for the QNMs; 
namely, use (\ref{eq:result}) to obtain $\omega_1$, and (\ref{eq:function}) 
to get $f_1$; this is then put into (\ref{eq:vn}) to find $V_2$, etc.  

\section{Explicit form of higher-order corrections and an 
example\label{sec:3}}

\subsection{Higher-order corrections}

The perturbative formulas would be more useful if they could be written
explicitly rather than recursively.  In general, the $n$th order correction 
to the frequency 
must take the form of an integral over $V(x_1) \cdots V(x_n)$; moreover, the 
perturbing potential can only act if it is ``sampled'' by the 
wavefunction $\phi_0^2(x)$.  It will also turn out to be convenient to remove
a constant from $V_1$, and we are led to define

\beq
W(x) = \left[V_1(x) - 2 \omega_0 \omega_1 \right] \phi_0^2(x)\;.
\eeql{eq:w}

\nid
The constant subtracted renders the integral of $W$ zero (see 
(\ref{eq:result})).

Thus we expect to be able to write the $n$th order correction in the form

\beq
\omega_n = \frac{1}{ 2\omega_0 \langle \phi_0 | \phi_0 \rangle} \,
\int dx_1 \cdots dx_n \, S \,
\prod _{j=1}^{n} W(x_j) \, \Psi_n(x_1, \cdots , x_n)\;,
\eeql{eq:expn}

\nid
where
$S \equiv \theta(x_1 - x_2) \cdots \theta(x_{n-1} - x_n)$
makes use of the symmetry among the coordinates $x_1, \cdots , x_n$ to restrict
the integration to one sector ($\theta$ is the unit step function), 
and the weight function $\Psi_n$, constructed out of
$\phi_0$, scales as $(\phi_0^2)^{1-n}$.

We now try to write out $\omega_2$ in essentially this form, and evaluate
the weight function $\Psi_2$.  For this purpose we consider the simpler case
of a half line $0 < x < \infty$, with the potentials satisfying $V_0(x) = V_1(x) = 0$
for $x > a$, and all $\phi(x=0)=0$. (This may be regarded as the odd-parity 
sector of a symmetric problem.) 
Thus, all the surface contributions at $L_-$ are
eliminated, while at the right hand side we may take $L_+ = a$ and 
$D_{+0}' = i$. Thus, the generalized norm is

\beq
\norm{\phi_0}
= \int_0^a dx\, \phi_0^2 (x) \,+\, \frac{i}{2\omega_0} \phi_0^2(a)\;.
\eeql{eq:norm3}

By using (\ref{eq:vn}) for $V_2$, we can write the second-order matrix 
element as\ftnote{2}{For QNMs, the wavefunction does not have any nodes
apart from the one at $x=0$ imposed by the boundary condition.}

\beq
\me{\phi_0}{V_2}{\phi_0}=-\int_0^a dx\, \phi_0^{-2}(x) \left[ f_1(x) 
\phi_0^2 (x) \right]^2- \omega_1^2 \int_0^a dx \phi_0^2(x)\;.
\eeql{eq:v2}

\nid
Now from (\ref{eq:function}) we have $f_1(x) \phi_0^2(x) = \int_0^x dy\, 
W(y)$. 
Putting this into (\ref{eq:v2}), changing the order of integration 
and also using (\ref{eq:norm3})
to simplify the second integral in (\ref{eq:v2}) then leads to

\beq
\fl \omega_2=\frac{1}{ 2\omega_0 \langle \phi_0 | \phi_0 \rangle} \,
\int dy dz\, S \, W(y) W(z) \Psi_2(y,z)
-\frac{\omega_1^2}{2\omega_0}+\frac{i\omega_1^2}{4\omega_0^2} 
\frac{\phi_0^2(a)}{\langle \phi_0 | \phi_0 \rangle }\;,
\eeql{eq:om2exp}

\nid
in which the weight function is found to be

\beq
\Psi_2(y,z) = \Psi_2(y) = -2 \int_y^a dx\, \phi_0^{-2}(x)\;.
\eeql{eq:f2}

\nid
Thus, except for the last two terms in (\ref{eq:om2exp}),
which do not involve an integral, the second-order correction
has been cast in the form (\ref{eq:expn}), which is the
most explicit form possible for an arbitrary perturbation.

\subsection{Example}

We now illustrate these formulas by a very simple example.  Let the
unperturbed system be defined by a step

\beq
V_0(x) = V_0 \, \theta(b-x) \;, \qquad b<a \;.
\eeql{eq:v0}

\nid
The unperturbed eigenfunctions are

\beq
\phi_0(x) = \left \{ \begin{array}{lll} 
 A \sin \! qx \; & , & ~~x \leq b \;,\\
 A \sin \! qb \, e^{i\omega_0 (x-b)} \; & , & ~~x>b \;,
\end{array} \right. 
\eeql{eq:phi0}

\nid
where the condition of outgoing waves results in the eigenvalue equation for 
$q$: 

\beq
q \cot \! qb = i\sqrt{ q^2 + V_0} \equiv i\omega_0\;.
\eeql{eq:q}

\nid
Here $\omega_0$ is the unperturbed frequency.  There are of course many solutions to (\ref{eq:q}),
and we pay attention to any one of these.

The norm is readily evaluated by (\ref{eq:norm3}) to be

\beq
\norm{\phi_0} =
\frac{A^2 b}{2} \left( 1 - \frac{\sin \! 2qb}{2qb}-\frac{\sin^2 \! qb
\, \tan \! qb}{qb} \right)\;.
\eeql{eq:norm4}

\nid 
For any perturbation $V_1$ with support on $(0,a)$, the first-order shift is 
then

\beq
\omega_1 = \frac{ \int_0^b dx \sin^2 \! qx \, V_1(x)\, + \, \int_a^b dx \, 
\sin^2 \! qb \, e^{2i\omega_0 (x-b)}\, V_1(x)}
{ \omega_0 b \left( 1 - \sin \! 2qb / 2qb -\sin^2 \! qb\, \tan \! qb/qb 
\right) }\;,
\eeql{eq:om1exm}

\nid
while the second-order correction is given by (\ref{eq:om2exp}), with
the weight function being

\beq
\fl \Psi_2(y) = \left \{ \begin{array}{lll}
(2/A^2)(\cot \! qb - \cot \! qy)+C
\left[1-e^{2i\omega_0 (b-a)}\right] & , & ~~y<b \;, \\
C\, (e^{-2i\omega_0 y}-e^{-2i\omega_0 a})& , & ~~y \geq b \;,
\end{array} \right.
\eeql{eq:w2}
where $C=i/(\omega_0 A^2 \, \sin^2 \! qb)$. 

These formulas then allow the corrections for any perturbation $V_1$ to be obtained
by direct quadrature; more importantly, they exhibit how the perturbation
$V_1$ acts to shift the complex eigenvalues. 

To be specific, let the perturbation be a bump of width $w$ centered at a 
position $x_0$: 

\beq
  V_1(x)=\left \{ \begin{array}{lll} 
  \displaystyle{1 \; }& , & 
\displaystyle{~~x_0-\bfrac{w}{2}<x<x_0+\bfrac{w}{2}} \;, \\
  \displaystyle{0 \; }& , & 
\displaystyle{\mbox{~~otherwise}} \;.\end{array} \right. 
\eeql{eq:vbump}

\nid 
Figure~1 shows the trajectory of the lowest eigenvalue, $\omega$, in the 
complex frequency plane as $x_0$ is changed, for fixed $V_0=100$, $b=1$, 
$\mu=10$, $w=0.1$ and (a) $V_1$ lying inside the interval $(0,b)$, (b) $V_1$ 
lying outside the interval $(0,b)$; the exact results (circles), 
first-order perturbation 
computed by (\ref{eq:om1exm}) (dashed line) and second-order 
perturbation computed by (\ref{eq:om2exp}) (solid line) are shown together 
for comparison. Figure~2 
shows the magnitude of the remaining error for the unperturbed eigenvalue 
(solid line), first-order perturbation (dashed line) and second-order 
perturbation (long broken line) versus $\mu$, for fixed $V_0=100$, $b=1$, 
$w=1$ and (a) $x_0=0.3$, (b) $x_0=1.4$. 

Although this example is extremely simple, it illustrates several interesting 
features. First, the remaining error of the $n$th-order perturbation scales as 
$\mu^{n+1}$, as expected. This is the case even for the perturbation lying 
outside the interval $(0,b)$ (Figure~2b), in which case the spectrum of QNMs 
is {\it not} complete, and it is not possible to write the second-order 
correction as a sum over intermediate QNMs; this result for the 
second-order correction is testimony to the utility of LPT (as opposed to 
RSPT). 

Comparison between the two cases in Figure~2a and 2b also reveals that 
higher-order corrections are more significant when the perturbation acts at 
a more distant position, where $|\phi_0|^2$ is large. 

The behavior in Figure~1 is even more interesting, showing a spiral 
structure as the position $x_0$ is changed. Although the perturbation is 
real and positive, the shift can have any phase depending on where the 
perturbation acts --- a situation totally different from NMs in conservative 
systems. This behavior is most readily understood in first-order perturbation 
theory, for which the general result (\ref{eq:result}) can be written as 

\beq
\frac{\delta \omega}{\delta V(x)}=H(x)\equiv \frac{\phi_0^2(x)}{\norm{\phi_0}}\;,
\eeql{eq:h}

\nid
where $H(x)$ can be read off from (\ref{eq:phi0}) and (\ref{eq:norm4}), which 
makes the spiral structure easy to understand. 
The pattern of the shifts can therefore be very rich, and an example of the 
results for a model astrophysical perturbation of a black hole has been
given recently \cite{dirt}. 

\section{Potentials with tails\label{sec:4}}

\subsection{The P\"{o}schl-Teller Potential} 

When dealing with QNMs, one recurring complexity is the
asymptotic behavior $\phi_0^2(x) \sim e^{2\gamma_0 |x|}$,  
making norms and matrix elements divergent.  As far as
LPT is concerned, this complication occurs at two different
levels.  If the potential (and its perturbation) vanishes outside
a finite domain, then the expression (\ref{eq:norm2})
suffices to produce a finite expression for the norm, whereas the matrix
elements involve integrals only over finite domains.
Such simplifications also extend to
potentials that vanish at infinity faster than any exponential.  However, 
when the potential (or its perturbation) decays as an exponential
or slower (which we shall refer to as a tail), then the evaluation of the 
norm and the matrix element will require more attention. 

In this Section we illustrate the solution of these problems
with the example of the P\"{o}schl-Teller (P-T)
potential \cite{po-teller}

\beq
V(x) = V_0 \cosh^{-2}(x/b)\;.
\eeql{eq:po-teller}

\nid
From the point of view of LPT, the P-T potential is interesting because its
large $|x|$ behavior is exactly exponential: $V(x) \propto e^{-2x/b}$.
The P-T potential, as one of a few exactly solvable models, has been
studied in depth, in part as a proxy for the Regge-Wheeler 
potential \cite{chand,bholen,others} or the Zerilli potential
\cite{zerilli}, which describes linearized gravitational waves 
propagating on a Schwarzschild background.  These also
have exponential tails (as the tortoise coordinate $x \rightarrow -\infty$,
i.e., towards the event horizon), and consequently
their QNMs share certain key properties with those of the P-T potential
(e.g., a string of QNMs evenly spaced ``vertically'' in the complex 
$\omega$-plane, 
$-\mbox{Im }\omega (j) \propto (j+\sfrac{1}{2})$\ftnote{3}{The mode index 
will be indicated in $( \, )$, but will be suppressed where no confusion 
arises.}).
Therefore a better understanding of the exponential tails
may be relevant to gravitational waves as well.

The QNM eigenvalues of (\ref{eq:po-teller}) are \cite{po-teller}

\beq
\omega (j) = 
\frac{1}{b} \left[ \pm \sqrt{V_0b^2 - \sfrac{1}{4}} -i(j+\sfrac{1}{2}) \right]\;,
\eeql{eq:eigenpt}

\nid
where we have assumed $4V_0b^2 > 1$.
The positive (negative) parity sector corresponds to even (odd) $j$.
For the purpose of illustrating the LPT formalism,
we shall focus on the lowest state in each sector, i.e., $j=0$ and $j=1$.

Consider perturbations of the width, specifically 

\beq
\frac{1}{b} = 1 + \mu\;.
\eeql{eq:perwidth}

\nid
Because the model is exactly solvable for all $b$, 
we immediately obtain the frequencies
in powers of $\mu$:

\bea
\omega_0 &=& \sqrt{V_0 - \sfrac{1}{4}} - (j+\sfrac{1}{2}) i 
\equiv \sigma - (j +\sfrac{1}{2})i\;,
\nn 
\omega_1 &=& \frac{1}{4\sigma}-(j+\sfrac{1}{2})i\;,
\eeal{eq:exeig}

\nid
etc., where for simplicity we only show the one eigenvalue of each pair 
with $\mbox{Re }\omega >0$.

We now show how the shift $\omega_1$ can be obtained from LPT.
From (\ref{eq:po-teller}) and (\ref{eq:perwidth}),

\bea
V_0(x) &=& V_0 \cosh^{-2} \!x \;,\nn
V_1(x) &=& -2 V_0 x \sinh \! x \cosh^{-3} \! x\;.
\eeal{eq:v01}

\nid
We show three different ways of handling the divergent integrals 
(\ref{eq:matelmf}) and (\ref{eq:normf}). The first 
two methods are specific to the P-T potential (or other potentials amenable 
to analytic treatment), but these lead to the third method, which is 
numerical and can be applied to any potential with exponential tails. The 
last method will be the one of general interest. 

\subsection{Analytic continuation}

Consider for example the $j=0$ state.  The unperturbed 
eigenfunction is\ftnote{4}{This expression applies only for $\omega$ equal 
to the eigenvalue; otherwise there is another term with an incoming wave, 
whose coefficient vanishes at the eigenvalue.  The omission of this term
does not affect the argument based on analytic continuation.}

\beq
\phi_0(x) = (\cosh x)^{i\omega} \; , \qquad \mbox{$j=0$}\;,
\eeql{eq:state0}

\nid
where it is understood that $\omega$ is to be evaluated at the
unperturbed value $\omega = \omega_0 = \sigma -i/2$.
Take the formal expression (\ref{eq:normf})
and define, for any $\omega$ for which the integral converges,

\beq
N(\omega) = \int_{-\infty}^{\infty} dx \phi_0^2(x)\;,
\eeql{eq:nom}

\nid
where $\phi_0$ is given by (\ref{eq:state0}).  The integral $N(\omega)$
is well-defined for $\mbox{Im } \omega > 0$, in which domain it is
evaluated in terms of the beta function $B$ to be

\beq
N(\omega) = B( \sfrac{1}{2} , -i\omega)\;.
\eeql{eq:normst0}

\nid
By analytic continuation, this applies to \mbox{$\mbox{Im } \omega < 0$}
as well, and the norm of the $j=0$ state is then
\mbox{$B(\sfrac{1}{2}, -\sfrac{1}{2}-i\sigma)$}.  Likewise, the wavefunction
for the $j=1$ state is

\beq
\phi_0(x) = \tanh \!x \, (\cosh \! x)^{i\omega} \; , \qquad \mbox{$j=1$}\;.
\eeql{eq:state1}

\nid
The same analytic continuation gives the norm as
\mbox{$B(\sfrac{3}{2}, -\sfrac{3}{2}-i\sigma)$}.

The matrix element $\me{\phi_0}{V_1}{\phi_0}$
for the $j=0$ state is

\beq
-2V_0 \int_{-\infty}^{\infty} dx\, x\sinh \! x (\cosh \! x)^{2i\omega-3}
= -\frac{\sqrt{\pi}V_0\, \Gamma(1-i\omega)}{(1-i\omega)\, 
\Gamma(\sfrac{3}{2}-i\omega)} \;,
\eeql{eq:mest0}

\nid
which is convergent even at $\omega = \omega_0 = \sigma -i/2$, and 
readily evaluated to be 
\[
  \me{\phi_0}{V_1}{\phi_0}=-\frac{\sqrt{\pi}V_0\, \Gamma(\sfrac{1}{2}-i\sigma)}
{(\sfrac{1}{2}-i\sigma)\, \Gamma(1-i\sigma)}\;.
\]
Thus the first-order shift is obtained from
(\ref{eq:normst0}) and (\ref{eq:mest0}) to be 
\[
 \omega_1=\frac{1}{4\sigma}-\frac{1}{2}i \;,
\]
in agreement with (\ref{eq:exeig}).

Similarly, for the $j=1$ state, the matrix element is

\bea
\fl -2V_0\int_{-\infty}^{\infty} dx\, x\sinh^3 \! x \, 
(\cosh \!x)^{2i\omega-5} \nonumber \\
\lo=\frac{V_0}{i\omega-2}\left[ B(\sfrac{3}{2},1-i\omega)  
+\bfrac{\pi \, 
\Gamma(1-i\omega)}{(1-i\omega)\, \Gamma(\sfrac{3}{2}-i\omega)}\right] \;,
\eeal{eq:mest1}

\nid
where the integral is evaluated for \mbox{${\rm Im}\, \omega \, >0$}. 
Analytic continuation to the eigenvalue is required, and gives 
\[
  \me{\phi_0}{V_1}{\phi_0}=\frac{V_0}{i\sigma-\sfrac{1}{2}}\left[ 
B(\sfrac{3}{2},-\sfrac{1}{2}-i\sigma)-\frac{\sqrt{\pi}\, 
\Gamma(-\sfrac{1}{2}-i\sigma)}{(\sfrac{1}{2}+i\sigma)\, \Gamma(-i\sigma)}
\right]\;.
\]
The first-order shift of the $j=1$ state is again in
agreement with (\ref{eq:exeig}).

Analytic continuation, though convenient,
applies only when the integrals can be evaluated
exactly.  We therefore present other methods, including
numerical evaluation of the integrals.

\subsection{Regularization\label{sec:4.3}}
The second method does not make use of the formal expression (\ref{eq:normf}),
but utilizes the original expression (\ref{eq:norm}) with the regulating 
parameters $-L_{-} = L_{+} = L$.  The integral involved is,
for the $j=0$ state,

\bea
\fl \int_{-L}^{L} dx\, \phi_0^2(x) &=& \int_0^{\tanh^2 \!L} dz\, z^{-1/2}
(1-z)^{-i\omega_0 -1} \nn
 &=& 2F(\sfrac{1}{2}, 1+i\omega_0; \sfrac{3}{2} ; \tanh^2 \!L )\, 
\tanh \!L \nn 
& = & B( \sfrac{1}{2} , -i\omega_0) 
-\frac{i\phi^2_0(L)}{\omega_0}
F(1, \sfrac{1}{2} -i\omega_0; 1-i\omega_0; \cosh^{-2}\!L)\, \tanh \! L \;,
\eeal{eq:normst0a}

\nid
where $F(a,b;c;x)$ is the hypergeometric function, and
the last step follows from its transformation properties.
This then gives

\beq
\norm{\phi_0}
= B(\sfrac{1}{2}, -i\omega_0) - \frac{\phi_0^2(L)}{ \omega_0} K(L)\;,
\eeql{eq:normst0b}

\nid
where

\beq
K(L)=D_{+}'-iF(1, \sfrac{1}{2}-i\omega_0; 1-i\omega_0; \cosh^{-2}\!L)
\, \tanh \!L \;.
\eeql{eq:normst0c}

\nid
Now it is easily shown, using even the crudest approximation
$D_{+} = i\omega_0$, that $K(L)=O(e^{-2L})$, whereas
$\phi_0^2(L) = O(e^L)$.  Thus the second term in (\ref{eq:normst0b})
(which is guaranteed to be independent of $L$) is shown to be zero
when evaluated at $L \rightarrow \infty$.  The norm of the $j=1$ state
can also be recovered in this manner, though one needs a better
approximation for $D_{+}$ in this case; a way to obtain these
better approximations is given below.

The matrix elements are likewise regulated, provided we know
the logarithmic derivatives $D_{\pm}(\omega)$ when the
wavefunction is integrated from $x \rightarrow \pm \infty$. These are not 
available for a general potential. Another way to regulate the matrix elements 
is to integrate along a contour in complex $x$ plane, e.g., along the path 
$x=u e^{i\theta}$, $u$ real and $\theta$ fixed. The matrix element becomes 

\beq
  \me{\phi_0}{V_0}{\phi_0}=\int_{-\infty}^{\infty} du\, e^{i\theta}
\phi_0^2(u e^{i\theta}) V_0(u e^{i\theta})\;.
\eeql{eq:me2}

\nid
It is easily seen that $\phi_0^2(u e^{i\theta})$ decays exponentially 
for sufficiently large $\theta$ and hence the integral along the rotated 
contour converges \cite{c}. 
As an example, we compute the matrix element numerically for $j=1$ state using 
(\ref{eq:me2}) with $\theta=60^{\circ}$. The result agrees with the analytic 
value given by (\ref{eq:mest1}).
In the case of P-T potential, the logarithmic derivatives $D_{\pm}(\omega)$ 
are available analytically. Hence the matrix element can also be evaluated 
by (\ref{eq:matelm}), but since this result is highly special, we shall not 
exhibit it here. Instead, we go on to a numerical scheme applicable to
all potentials with exponential tails.

\subsection{Numerical evaluation and Born series}

For any system, provided the unperturbed wavefunction $\phi_0(x)$
is known, the shifts are in principle given by (\ref{eq:matelm})
and (\ref{eq:norm}).  These involve (a) finite integrals from
$L_{-}=-L$ to $L_{+}=L$, which can be handled numerically
in the usual way; and (b) surface contributions involving $D_{\pm}$.
The latter contain all the information from the tails of the potential.
The lowest approximation
$D_{\pm}(\omega) \approx \pm i\omega$
is in general not accurate enough, because
it multiplies $\phi_0^2(L) \sim e^{2\gamma_0 L}$.

A general, yet simple way to obtain a better approximation for $|x|>L$ 
is to use the Born approximation.
Here $V$ will stand for {\em any} potential;
by applying the method
sketched below, we can find the logarithmic derivative
of either $V_0$ or $V_0 + V_1$, and hence obtain the
quantities $D_{\pm}$.
If $V(x) \propto e^{-\alpha x}$ and 
the Born approximation
is carried to $m$th order, then the remaining error would go as 
$V(x)^{m+1} \propto e^{-(m+1)\alpha x}$, which will
be sufficiently accurate for dealing with any unperturbed
state with $\gamma_0 < (m+1)\alpha$.

The Born approximation is particularly easy to implement for a potential
that goes as an exponential. For simplicity we deal with the tail at $x=L$ 
only. Let

\beq
V(x) = V_0 \, \sum_k c_k e^{-\alpha_k x} \;,
\eeql{eq:sumexp}

\nid
where by convention $\alpha_1 < \alpha_2 < \cdots$.  The 
P-T potential is of this form, where
$\alpha_k = 2k/b$, and $c_k=(-1)^{k+1} 4k$.

By iterating (\ref{eq:ricc}) in powers of $V$, one finds

\bea
\fl f_0(x) &=& i\omega \;,\nn
\fl f_1(x) & = & V_0 \, \sum_k 
c_k (\alpha_k - 2i\omega)^{-1} e^{-\alpha_k x} \;,\nn
\fl f_2(x) & = & V_o \, \sum_{k,k'} c_k c_{k'} 
(\alpha_k -2i\omega)^{-1} (\alpha_{k'} -2i\omega)^{-1}
(\alpha_k + \alpha_{k'} -2i\omega)^{-1}
e^{-(\alpha_k + \alpha_{k'} )x}\;.
\eeal{eq:born}

\nid
etc. (Here the subscripts on $f$ denote the order of the Born
approximation, not LPT.)
Higher-order terms can be generated readily by algebraic software.  
All the sums can
be terminated at some $k_{max}$ if only accuracy up to 
$O(e^{-\beta x})$ is required for some finite $\beta$.

The poles in (\ref{eq:born}) exist only in the Born approximation, but not
in the exact solution~\ftnote{5}{In deriving the Born approximation,
one has in effect first taken $V_0 \rightarrow 0$, then secondly
considered say $\alpha_k-2i\omega \rightarrow 0$ in the resultant
expression (\ref{eq:born}).  This order of the limits implies
that the result is only valid for
$V_0 e^{-\alpha_k x} \ll | \alpha_k -2i\omega |$.  On the other hand,
the exact solution at the position in question would refer to taking the
limit $\alpha_k-2i\omega \rightarrow 0$ while keeping $V_0$ finite.  In
this case, there would be no pole.}.
Nevertheless, (\ref{eq:born}) makes it clear that even exponentially small
potentials can have a significant effect when $-\mbox{Im } \omega$ is large, 
and this is the reason behind the string of QNMs 
\mbox{$-$Im $4M\omega$}$\approx (k+\sfrac{1}{2})$ for a Schwarzschild black 
hole of mass $M$ \cite{highdamp}. 

We have implemented this scheme for the P-T potential, indeed for
any potential that can be expressed in the form (\ref{eq:sumexp}) with
$\alpha_k = k \alpha$.  Again for simplicity we deal with
the situation only on one side, say for the tail as $x \rightarrow +\infty$. 
For this particular form of $\alpha_k$ spaced evenly in $k$, the Klein-Gordon 
equation can be solved easily by substituting 
\beq
  \phi(x)=e^{i\omega x}\sum_{k=0}^{\infty} d_k e^{-k\alpha x}
\eeq
into (\ref{eq:kgom}). One has 

\bea
 d_0 &=& 1 \;,\cr
 d_k &=& \bfrac{V_0}{\alpha k(\alpha k-2i\omega)}\sum_{m=0}^{k-1}d_m c_{k-m}
\;, \qquad k\geq 1 \;,
\eeal{eq:d}
and

\beq
f(x) = \left[\sum_{k=0}^{\infty}(i\omega-\alpha k)d_k e^{-\alpha kx}\right]
\left(\sum_{k=0}^{\infty}d_k e^{-\alpha kx}\right)^{-1}\;.
\eeql{eq:logderiv2}

\nid
In this example, we have summed four terms with $\alpha=2/b=2$, and consequently 
the remaining error in the logarithmic derivative calculated is $O(e^{-10x})$. 
Incidentally, this method, when evaluated at $x=L$, gives accurate expression 
for $D_+(\omega)$ as needed in \mbox{Section \ref{sec:4.3}}.

We have used this method to evaluate both the matrix element (\ref{eq:matelm}) 
and the norm (\ref{eq:norm}), taking $L_+=L=5$. (In this example, only the 
positive half line is needed due to the symmetry of the potential.) 
To be precise, the integral over the finite domain $(0,L)$ is evaluated 
numerically, while the surface term is evaluated by the Born series 
(\ref{eq:logderiv2}). The 
result for $\omega_1$ agrees accurately with the result obtained from the 
two methods sketched earlier. 

However, there is still a numerical problem. Take the norm in the $j=1$ state 
as an example. Numerical evaluation gives for the two terms in 
(\ref{eq:matelm}) 
\begin{eqnarray*}
  \mbox{integral} &=& 187374.578 + 143350.152 i \;,\\
  \mbox{surface term} &=& -187374.961 - 143350.431 i \;,
\end{eqnarray*}
so that there is a loss of 6 significant digits when the two terms are 
combined. The cause of the problem, as before, is the exponential growth 
of the wavefunction $\phi_0(x) \approx A e^{i\omega_0 x}$, so 
that the asymptotic $L$ dependence of the two terms are respectively 
$\pm (A^2/2i\omega_0)\, e^{2i\omega_0 L}$, where $A=1/2^{i\omega_0}$ for 
the wavefunction $\phi_0(x)$ normalized as in (\ref{eq:state1}). 
A related difficulty is that the integrand is large and oscillating, which 
limits the accuracy of evaluating the integral. However, these difficulties 
are readily remedied if we subtract $A^2 e^{2i\omega_0 x}$ from the integrand, 
and add the corresponding term $(A^2/2i\omega_0)\, (e^{2i\omega_0 L}-1)$ to 
the surface term. Then in this example one finds
\begin{eqnarray*}
  \mbox{modified integral} &=& -22.9370946 + 29.1215523 i \;,\\
  \mbox{modified surface term} &=& 22.5541253 - 29.4010213 i\;,
\end{eqnarray*}
and there is only a loss of 2 significant digits when the two terms are 
combined. This technique can be further refined by removing subasymptotic 
terms as well. This method does not rely on any property of the P-T 
potential other than the exponential tails. 

The numerical difficulty associated with the exponential growth of the QNM 
wavefunction is exactly the same as the difficulty in the ``shooting'' 
algorithm discussed in \mbox{Section \ref{sec:1.2}}. This same difficulty, in 
different guises, always besets numerical solutions of QNM problems. Here we 
have developed an effective method within the realm of perturbation theory 
--- but otherwise applicable to {\it any} system with exponential tails 
--- to tame the problem. 
The class of problems with exponential tails is sufficiently wide for this 
method to be of interest, especially since there is a dearth of other 
effective methods. With this numerical technique to handle exponential 
tails, LPT is completely formulated for potentials either without tails, 
or with exponential tails. 

\section{Conclusion\label{sec:5}}

In this paper we have formulated LPT for QNMs.
For systems without tails, the formalism is no more complicated
than for NMs.  In fact, there are several advantages: the absence of
nodes allows simple application to {\em all} states, not just
the ground state, and the possibility that QNMs may not be complete
makes alternative methods (e.g., generalization of RSPT) less
useful.  The explicit form of the first-order shift is given, as well
as the most general form of the second-order shift for an arbitrary 
perturbation.  When there is a tail 
that can be expressed as a sum of exponentials, a method is developed, 
based on the
Born series, that reduces the calculation to the evaluation of integrals, 
the exponentially large nature of which can be handled by subtracting off the 
leading asymptotic terms.  While this is somewhat involved, it is to be 
stressed that for this case {\em no} other methods apply in 
general, not even brute-force numerical integration, on account of the need to
extract an exponentially small ``wrong'' solution.  Thus the technique is
likely to be useful. Indeed, this technique has already been employed to 
deal with model perturbation of a black hole \cite{dirt}. 

Finally, the generalized norm plays a central role, and in fact has a 
significance beyond perturbation theory.  It emerges naturally
in the derivation, where the integrating factor in (\ref{eq:int})
is $\phi_0^2$ and not $|\phi_0|^2$.  However, it is possible to
express this same idea in another way, which is possibly more
natural and familiar \cite{sun}.  The idea is to write these
open systems in terms of a non-hermitian Hamiltonian \cite{twocomp1,twocomp2}
and adopt a bi-orthogonal
basis \cite{biorth} which includes a set of left-eigenfunctions
${\overline \phi}$ dual to 
the right-eigenfunctions $\phi$.  Then our generalized
norm $\norm{\phi}$ is exactly the same as $({\overline \phi}, \phi)$,
where the latter is the {\em conventional} inner product
which is conjugate linear in the bra and linear in
the ket.  This development will be reported elsewhere \cite{sun2}. 

\ack 
This work is supported in part by a grant from the Hong Kong Research
Grants Council under grant 452/95P. We thank C K Au for many discussions.


\newpage
\centerline{\epsfig{file=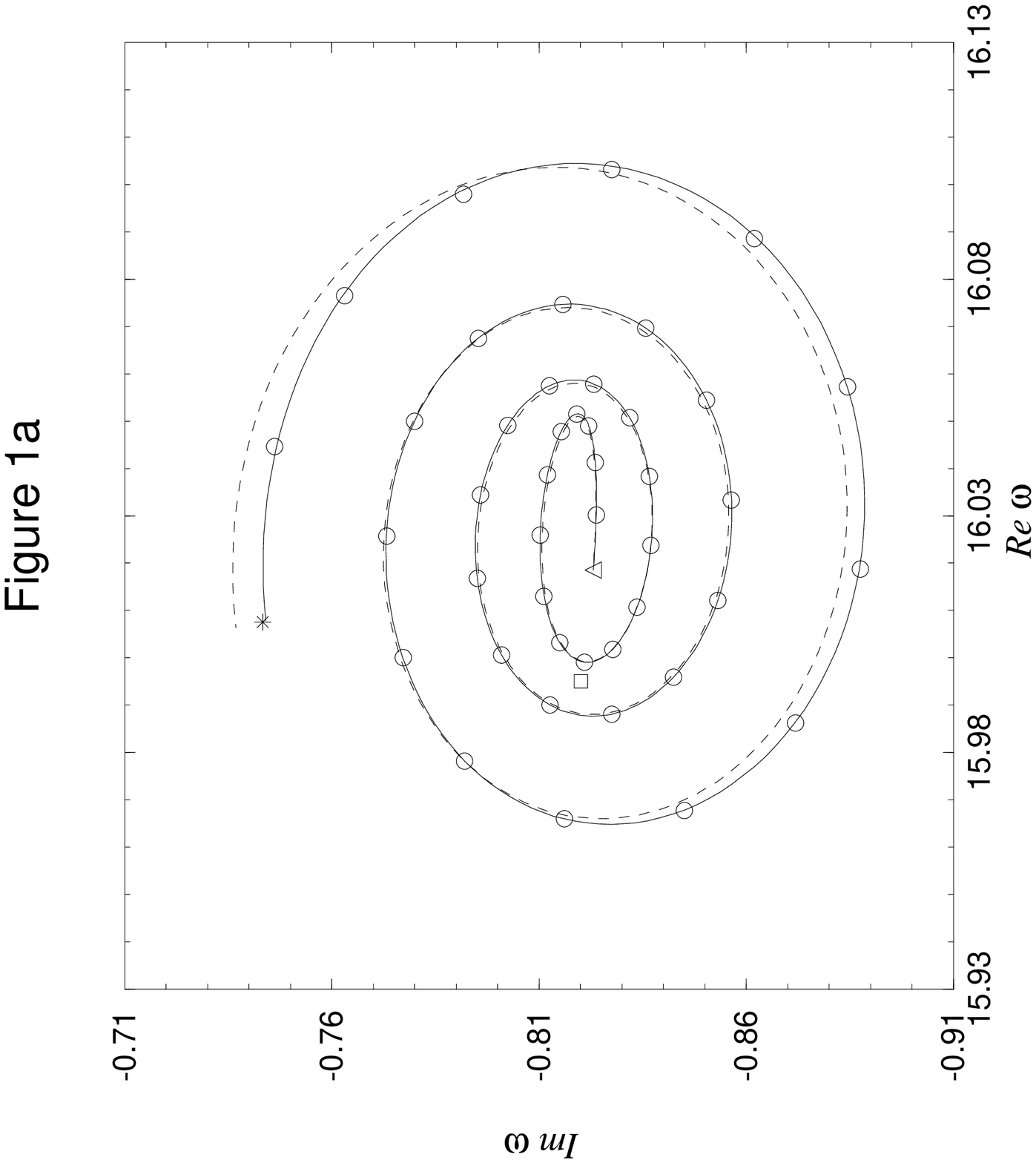,height=18cm}}

\centerline{\epsfig{file=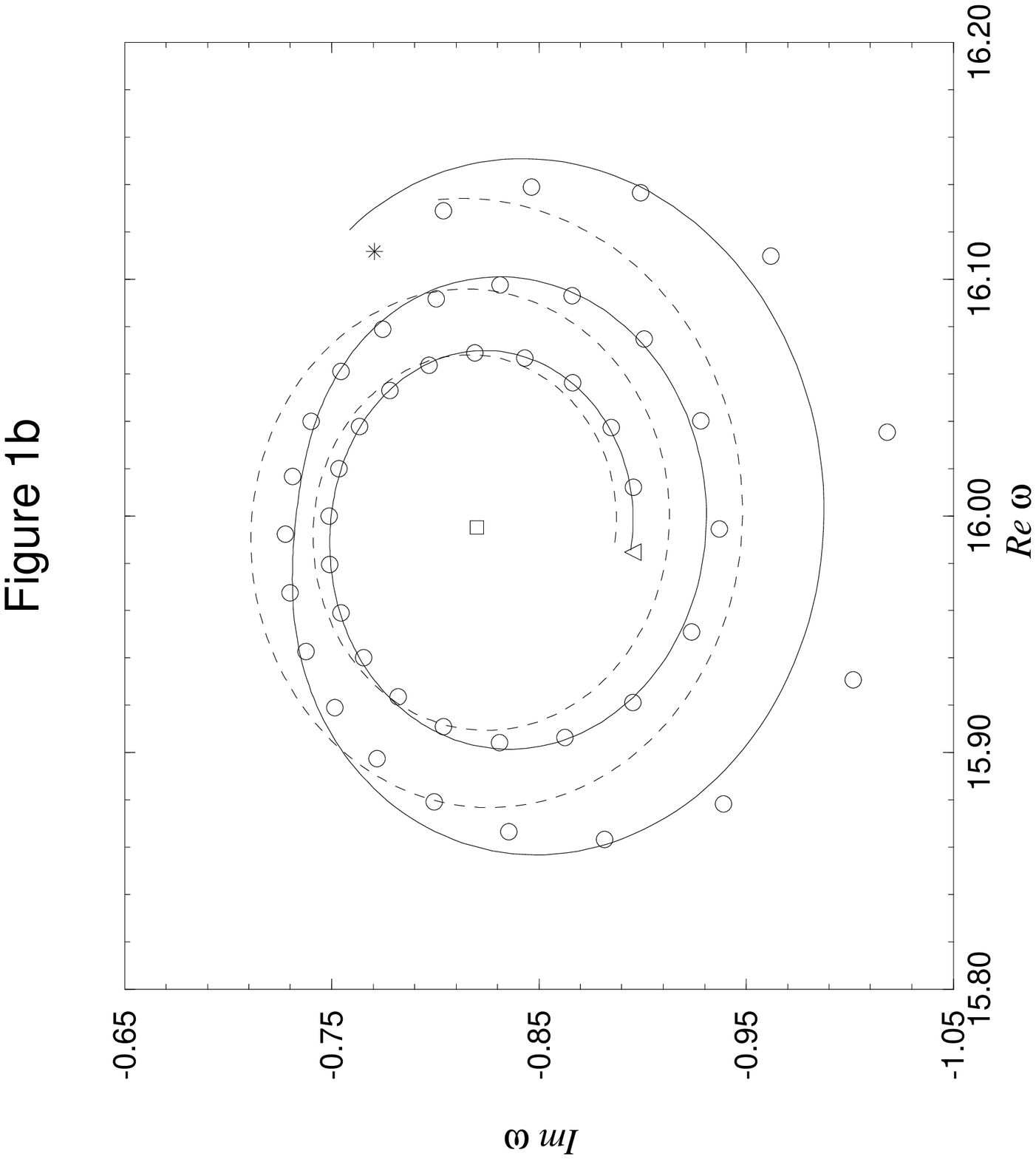,height=18cm}}

\centerline{\epsfig{file=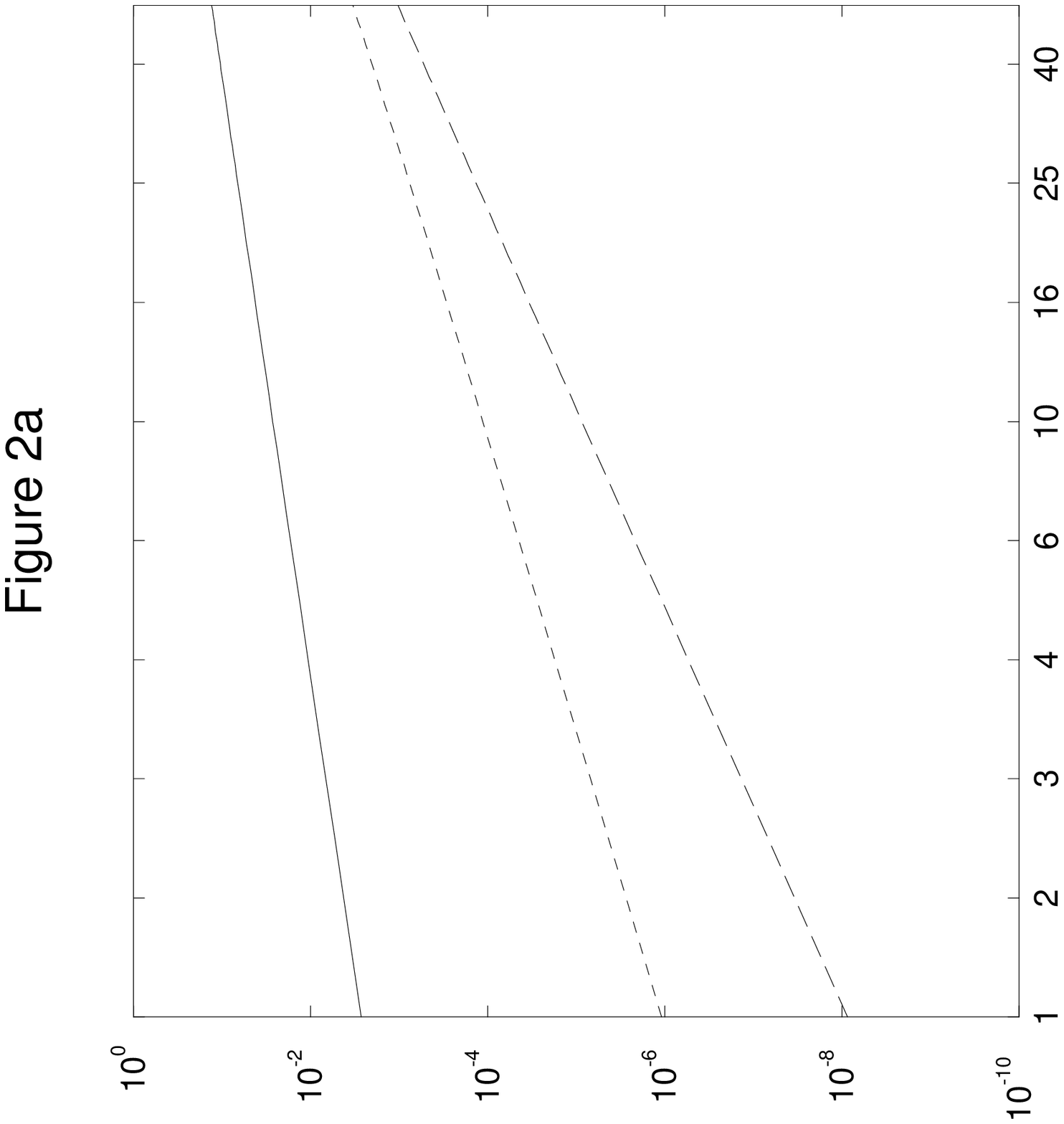,height=18cm}}

\centerline{\epsfig{file=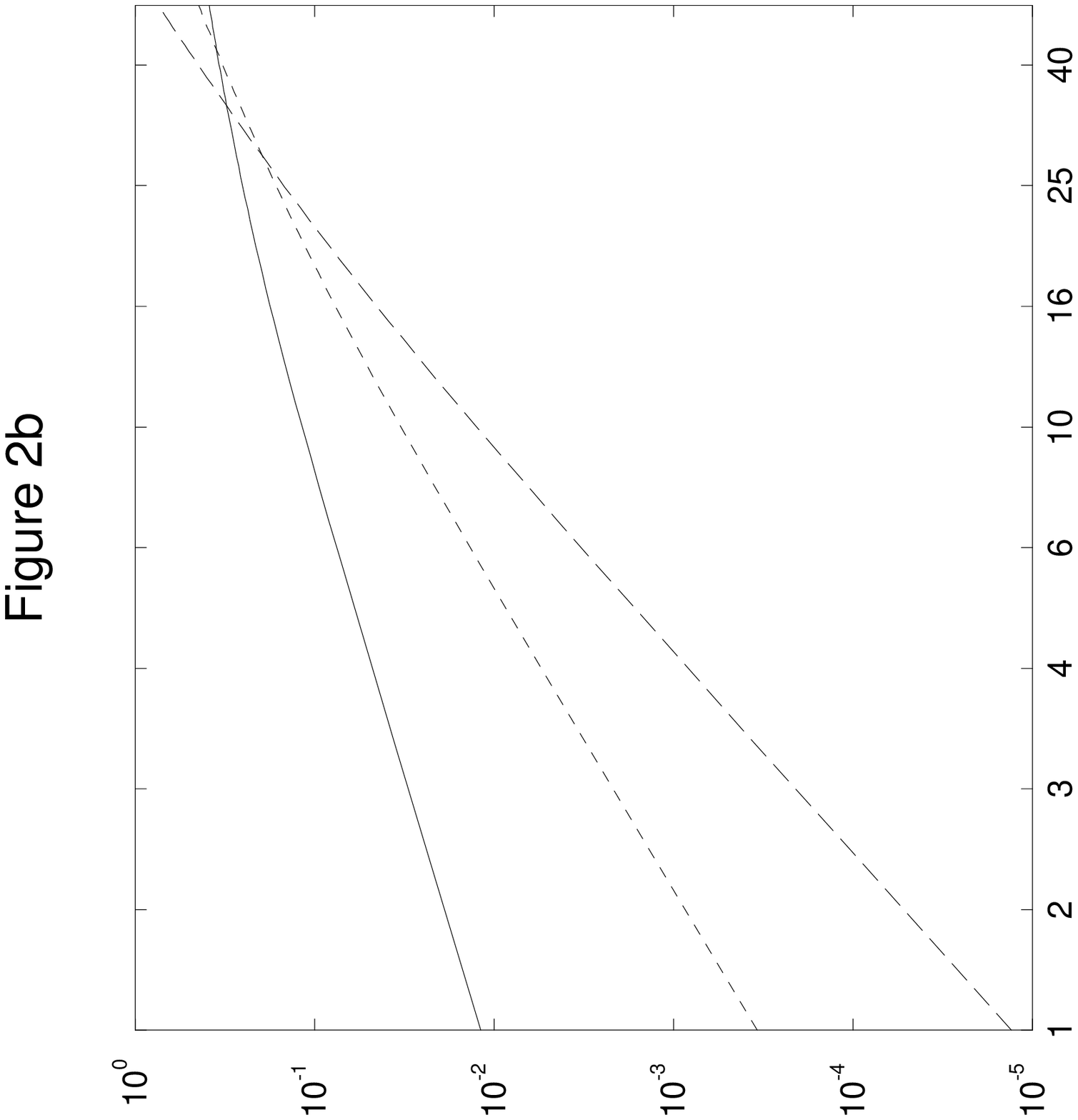,height=18cm}}

\Figures 

\nid
Figure~1a: The trajectory of the lowest eigenvalue in the complex 
frequency 
plane as $x_0$ is changed from $0.05$ to $0.95$, for fixed $V_0=100$, $b=1$, 
$\mu=10$, $w=0.1$. The first-order and second-order perturbation are 
indicated by the dashed line and the solid line respectively. The triangle 
and the star mark the 
positions of the exact eigenvalues for $x_0=0.05$ and 0.95 respectively. 
The circles show the positions of the exact results for other values of $x_0$, 
and the unperturbed eigenvalue is denoted by the square. 

\vskip 5mm
\nid
Figure~1b: Same as Figure~1a but for $1.05 \leq x_0 \leq 1.50$. The triangle 
and the star mark the positions of the exact eigenvalues for $x_0=1.05$ and 
1.50 respectively. 

\vskip 5mm
\nid
Figure~2a: The magnitude of the remaining error for the unperturbed value 
(solid line), first-order perturbation (dashed line) and second-order 
perturbation (long broken line) versus $\mu$, for fixed $V_0=100$, $b=1$, 
$w=0.1$ and $x_0=0.3$. 

\vskip 5mm
\nid
Figure~2b: Same as Figure~2a but with $x_0=1.4$. 

\end{document}